\documentclass[preprint,trackchanges]{aastex701}
\graphicspath{{./}{figures/}}
\usepackage{float}  
\usepackage{amsmath} 
\usepackage[dvipsnames]{xcolor}
\usepackage{subcaption}
\usepackage{xcolor} 
\usepackage{multirow}
\usepackage{makecell} 
\begin{document}

\title{Improving Solar Flare Soft X-ray Classification With FOXES: A Framework For Operational X-ray Emission Synthesis}

\author[orcid=0000-0003-3493-9174,sname='Goodwin']{Griffin T. Goodwin}
\affiliation{Physics \& Astronomy Department, Georgia State University, Atlanta, GA, USA}
\email[show]{ggoodwin5@gsu.edu}  

\author[orcid=0009-0000-3643-8061,sname='March']{Alison J. March}
\affiliation{College of Engineering \& Applied Science, University of Colorado at Boulder, Boulder, CO, USA}
\email[show]{alma2157@colorado.edu}

\author[]{Jayant Biradar}
\affiliation{College of Information Science, University of Arizona, Tucson, AZ, USA}
\email[show]{jayantbiradar@arizona.edu}

\author[orcid=0009-0004-2182-2596,sname='Schirninger']{Christoph Schirninger}
\affiliation{Institute of Physics, University of Graz, Graz, Austria}
\email[show]{christoph.schirninger@uni-graz.at}

\author[orcid=0000-0002-9309-2981,sname='Jarolim']{Robert Jarolim}
\affiliation{High Altitude Observatory,
  NSF National Center for Atmospheric Research,
  Boulder, CO, USA }
\email[show]{rjarolim@ucar.edu}

\author[orcid=0000-0002-8164-5948, sname='Vourlidas']{Angelos Vourlidas}
\affiliation{The Johns Hopkins University Applied Physics Laboratory, Laurel, MD, USA}
\email{}

\author[orcid=0000-0002-4001-1295, sname='Sadykov']{Viacheslav M. Sadykov}
\affiliation{Physics \& Astronomy Department, Georgia State University, Atlanta, GA, USA}
\email{}

\author[]{Lorien Pratt}
\affiliation{Quantellia LLC, Denver, CO, USA}
\email{}







\begin{abstract}
The Geostationary Operational Environmental Satellite (GOES) solar soft X-ray (SXR) irradiance in the 1--8$\rm \AA$ wavelength range is a long-standing measure of solar activity, used to define the classification of flare strengths. As a result, the flare class, along with the SXR light curves, are routinely used as a primary input for forecasting properties of space weather drivers, from coronal mass ejection speed to energetic particle output. However, the GOES SXR irradiance lacks spatial information, leading to known classification errors, such as misattributed flare locations during periods of high activity. Moreover, GOES only provides observations from Earth's orbit, hindering forecasting for other places in the heliosphere. Motivated by these limitations, we introduce the Framework for Operational X-ray Emission Synthesis (\emph{FOXES}), a Vision Transformer-based approach for translating Extreme Ultraviolet (EUV) spatially-resolved observations into SXR irradiance predictions. The model produces two outputs: (1) a global 1--8$\rm \AA$ SXR flux prediction and (2) per-patch flux contributions, which offer a spatially-resolved interpretation of where the model attributes SXR emission. Trained, validated, and tested on over 3200 hours of observations, \emph{FOXES} has demonstrated a translational mean absolute error of 0.051 dex for integrated SXR measurements. \emph{FOXES} has also shown promise in dissecting the solar background SXR flux during flaring and non-flaring events. Overall, this model paves the way for EUV-based spatially-resolved flare detection to be extended beyond Earth's line of sight. Such capabilities could lead to a more comprehensive flare catalog and enable a true multiviewpoint monitoring of solar activity.
 

 
\end{abstract}



\section{Introduction} 
\label{sec:intro}
Solar flares are radiative manifestations of rapid magnetic energy release in the solar atmosphere \citep{priest2002}. Although solar flares have mostly short-lived, recoverable impacts on Earth, their activity serves as one of the best predictors for larger events that can do long-lasting damage, such as coronal mass ejections and solar energetic particle events. Together, these phenomena have the potential to affect much of our core infrastructure, disrupting satellites, radio communications, GPS, and power grids once they reach Earth \citep{marusek_solar_2007,Schrijver2009, natras2019strong}. Consequently, it has become a high priority to develop accurate and reliable flare forecasting models to mitigate these risks \citep{barnes2016flare_workshop, Leka2019forecasting_methods}. In recent years, many teams have adopted data driven, statistical, and machine learning (ML) methodologies to address this problem, using various observational data products such as photospheric magnetograms, extreme ultraviolet (EUV) imagery, sunspot counts, and individual active region properties \citep{bobra_helioseismic_2014,huang_deep_2018,nishizuka_deep_2018,camporeale_challenge_2019,georgoulis_flare_2021, leka_properties_2023, ramos2023ml_solar,goodwin_investigating_2024}. Training such ML models, and generating statistically robust results, typically requires very large datasets, and given the terabyte-scale data streams now routinely delivered by space-based observatories, such as the Solar Dynamics Observatory (SDO; \citealp{Pesnell2012}), data volume is, in principle, no longer a limiting factor --- at least for a restricted subset of data products, such as EUV imagery and line-of-sight photospheric magnetograms. However, a critical prerequisite for the success of these approaches is the availability of \emph{comprehensive} and \emph{accurately} labeled flare catalogs. As is often noted in the community, the quality of training data fundamentally constrains the quality of the resulting models (i.e., ``garbage in, garbage out''). In the case of flare forecasting, if we have unreliable labels for the strength of flares, then our models will learn an inaccurate mapping between, say, magnetic field data and the probability of such an event happening within a certain prediction time window (see \citealp{wen_outlier_2025} as a case study).

The primary approach for the classification of solar flares is through the spatially integrated soft X-ray (SXR) irradiance from the Sun in the 1--8$\rm \AA$ wavelength range, measured by the Geostationary Operational Environmental Satellite (GOES) series \citep{woods2024}. Flares are categorized on a logarithmic scale provided by the National Oceanic and Atmospheric Administration (NOAA). The classes from weakest to strongest are as follows: \emph{A} ($< 10^{-7}\,W/m^2$), \emph{B} ($10^{-7}-10^{-6}\,W/m^2$), \emph{C} ($10^{-6}-10^{-5}\,W/m^2$), \emph{M} ($10^{-5}-10^{-4}\,W/m^2$), and \emph{X} ($\geq 10^{-4}\,W/m^2$). Since the SXR flux measured by GOES is the summed emission over the entire solar disk, and because no operational SXR imager with the necessary cadence and wavelength coverage exists or is planned for the near future, the spatial localization and disambiguation of individual solar flares is severely limited. 

Currently, there are several active catalogs maintained and updated on a daily basis, including those produced by the Space Weather Prediction Center (specifically their daily \texttt{Space Weather Event Reports})\footnote{\href{https://www.ngdc.noaa.gov/stp/space-weather/swpc-products/daily_reports/space_weather_event_reports/}{https://www.ngdc.noaa.gov/stp/space-weather/swpc-products/daily\_reports/space\_weather\_event\_reports/}}, Hinode\footnote{\href{https://hinode.isee.nagoya-u.ac.jp/flare_catalogue/}{https://hinode.isee.nagoya-u.ac.jp/flare\_catalogue/}}, and the Heliophysics Events Knowledgebase (HEK)\footnote{\href{https://www.lmsal.com/hek/}{https://www.lmsal.com/hek/}}. However, these catalogs exhibit a number of restrictions due to the limitations of GOES, such as temporal gaps, inaccuracies in reported flare locations, and ambiguities in the assigned flare classes when multiple events occur concurrently \citep{angryk2020flare_catalog}. Systems like HEK attempt to resolve this by exploiting Solar Dynamics Observatory Atmospheric Imaging Assembly (SDO/AIA; \citealp{Lemen2011}) EUV data to assist with flare localization; however, these systems primarily identify event locations through simple intensity enhancements at specific EUV channels and, most importantly, do not classify flares according to their SXR-based strengths. As a result, numerous efforts to refine and expand existing catalogs, in addition to developing methods for spatially resolving GOES SXR observations, have been done --- each playing a critical role in advancing the field (see, e.g., \citealp{padial_doble_alexis_2025, berretti_asr_2025}). In this work, we build upon these prior contributions by formulating the problem from a machine learning perspective, offering a complementary framework that facilitates further validation and assessment of existing catalogs. 

EUV images are among the most common data products collected by the current fleet of Heliophysics-focused spacecraft, with both Sun-Earth-line (e.g. SDO/AIA and GOES/SUVI) and off-Sun-Earth-line observations (e.g. STEREO/EUVI and Solar Orbiter/EUI) being widely available. These measurements allow us, in principle, to capture both the dynamics of the solar corona and the flaring activity of solar active regions. Given this fact and the abundance of EUV data, a natural question arises: can EUV observations alone be used to infer a flare's SXR emission? If so, this opens up the possibility of measuring flare strengths on board any of the available EUV imagers, paving the way for off-Sun-Earth-line SXR observations and related flaring properties. While there have been investigations into these translations (see \citealp{nitta2013,chertok_simple_2015,van2022,padial_doble_alexis_2025}), to the extent of our knowledge, no prior work has applied deep learning directly to full-disk, multi-channel EUV imagery to learn this mapping end-to-end.

Thus, we present \emph{FOXES: A Framework for Operational X-ray Emission Synthesis}. Utilizing Vision Transformers (ViTs; \citealp{dosovitskiy2020image}), \emph{FOXES} learns the mapping between EUV observations taken by SDO/AIA and the integrated SXR flux measurements of GOES. Then, as a consequence of our model's architecture, the integrated value can be spatially distributed across the solar disk, allowing for identification of source regions and their contributions to the estimated total flux, even during sympathetic flare events. This essentially gives us a virtual spatially-resolved GOES X-ray Sensor (XRS; \citealp{chamberlin2009}) on board of SDO, fusing the capabilities of the two instruments, and opening up the possibility to improve flare classification and localization.

The remainder of this paper is organized as follows: In Section \ref{sec:data}, we give a brief overview of the data used in this work, along with the steps employed to ensure that our data is ready for machine learning applications. In Section \ref{sec:methodology}, we provide a detailed description of our methodology, giving background for how we designed \emph{FOXES}. Finally, in Sections \ref{sec:results} and \ref{sec:conclusions} we provide results for our model, a case study highlighting \textit{FOXES'} potential implications for the future of flare forecasting, and our key takeaways.

\section{Data Collection and Preprocessing}
\label{sec:data}

\emph{FOXES} utilizes EUV observations from SDO/AIA as input to spatially map the disk-integrated SXR flux measured by GOES onto the solar disk. For this purpose, we use the seven AIA EUV channels that are particularly sensitive to hot flaring plasma: 94\AA, 131\AA, 171\AA, 193\AA, 211\AA, 304\AA, and 335\AA~\citep{petkaki2012sdo}. For the GOES SXR observations, we use the 1-minute averaged 1–8\AA~band flux, which is the standard for solar flare classification. Both the EUV and SXR datasets were obtained via the Joint Science Operations Center (JSOC) and through Fido using the \texttt{sunpy} \citep{sunpy2020} Python package. The temporal coverage was designed to maximize the number of flaring events in the sample.

Our data acquisition strategy proceeded as follows. First, using the GOES flare catalog, we identified flaring events (with peak magnitude of $\geq$M-class) that occurred between January 2012 and September 2025. For each event, we downloaded all EUV and SXR (\texttt{Release Version 2-2-0}) data at a 1-minute cadence from 15 minutes before the event start time to 15 minutes after the event end time (as defined in the SWPC GOES flare catalog). Subsequently, we adjusted the cadence of additional observations between successive events on the basis of their temporal separation. When the end time of one flare occurred within an hour of the start time of the next, we downloaded the intervening data at a 10-minute cadence. If the separation between events was less than one day, we used a 1-hour cadence; if the separation was less than 30 days, a 6-hour cadence was adopted. For separations exceeding 30 days, we downloaded the intervening data at a 1-day cadence. \emph{This enables us to capturing flare quiet data in addition to flaring events within our dataset.} To obtain more continuous data, we downloaded all available 1-minute cadence data between July 1st, 2023, and August 31st, 2023. This time range was selected as it will be particularly useful for validating future studies mentioned in Section \ref{sec:potential_implications}.

To obtain a machine learning-ready dataset, several steps were taken to preprocess both the AIA and SXR data. For the AIA data, we utilize the Instrument-to-Instrument (ITI; \citealp{jarolim2025}) tool for preprocessing. This pipeline includes a crop of the full Sun observations to 1.1 solar radii, a correction for instrument degradation (using ML; \citealp{dos_santos_multichannel_2021}), an image and exposure normalization, as well as resampling of images to a resolution of 512$\times$512 pixels (using a 4th order affine transformation). For GOES data, instruments with overlapping observations (e.g., GOES 13, 14, and 15, for the time period between 2015 and 2020) were averaged between available data points. Additionally, to obtain true fluxes for the GOES 13--15 operational data, the SWPC scaling factors were removed from the data. To do this, we divided the 1--8\AA~band flux for GOES 13--15 by 0.7 \citep{machol_goes_nodate}. This ensured proper scaling across instruments. Finally, for both GOES and AIA, only science-quality data were selected. For example, GOES data with eclipses and AIA data with artifacts were removed. This was done through monitoring quality flags when downloading the data, and ensures that we have a full stack of EUV data for a given time.

\begin{figure}[p]
  \centering
\includegraphics[width=1\textwidth]{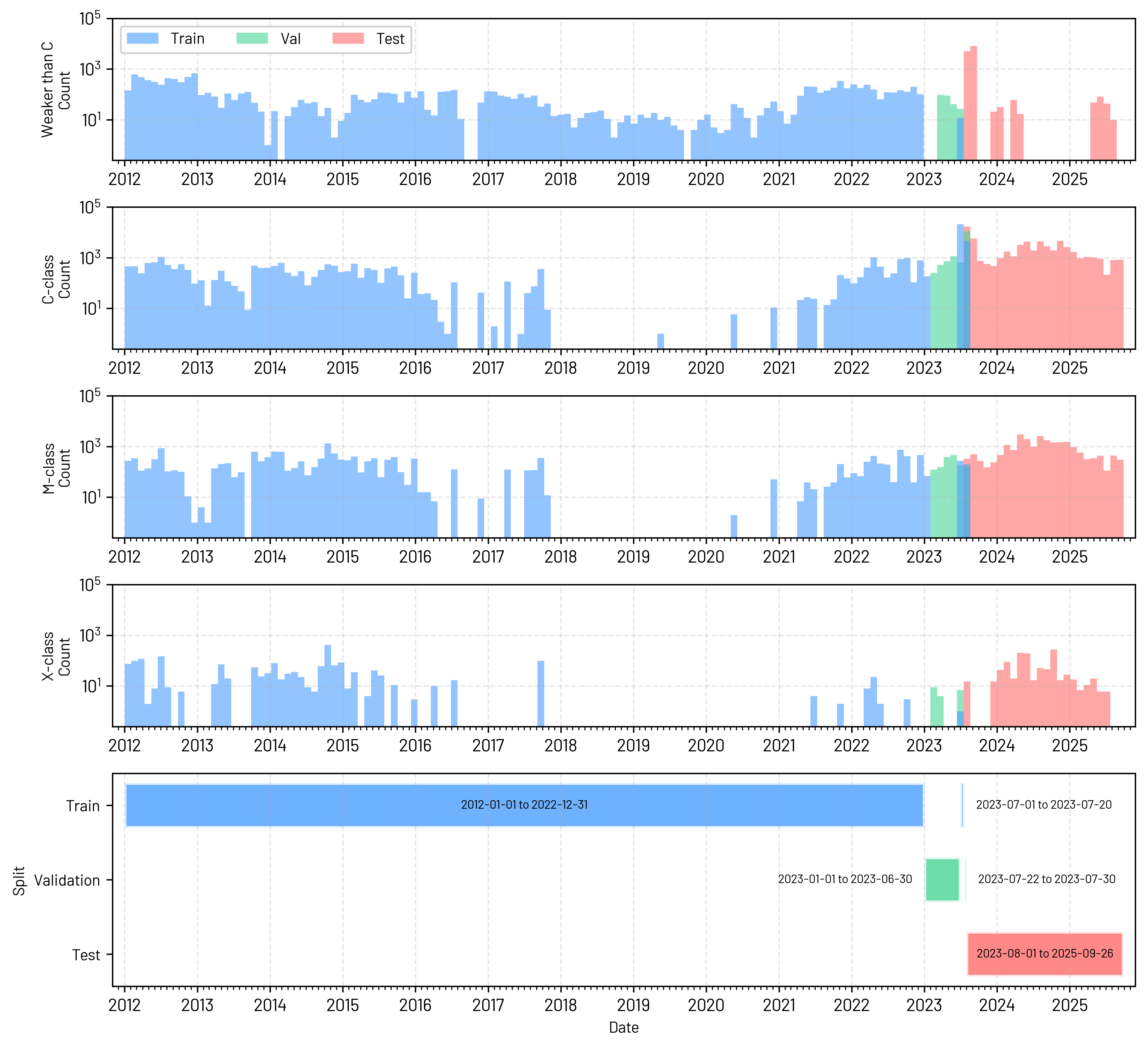}
  \caption{An overview of the data splits for training, validation, and testing. The top four panels show the breakdown of the number of data points for a given SXR level as a function of time. The colors illustrate the split the data belongs to: training (blue), validation (green), testing (red). The last panel shows the timeline for each split. \emph{Please note that even though there are overlapping bars between the data splits, this overlap is not actually present in the data and is just a consequence of the number of bars we are using in the visualization.}}
  \label{fig:datasplits}
\end{figure}

Once the data were downloaded and preprocessed, each stack of EUV images for a given point of time was directly matched with an integrated SXR value. The data was then divided into training, validation, and test splits, ensuring no temporal overlap between the data. An overview of the splits, along with the number of observations in each flare class, is provided in Table \ref{table:data} and Figure \ref{fig:datasplits}. In total, our dataset covers about 3,200 hours of observations and adds up to roughly 1.4 terabytes. Please visit HuggingFace\footnote{See\dataset[doi: 10.57967/hf/7737]{https://huggingface.co/datasets/griffingoodwin04/FOXES-Data} for access to the data.} to access the data \citep{griffin_goodwin_2026}.

\begin{table}[h]
\centering
\caption{Summary of the number of observations in the training, validation, and test sets for the different flare classes. The dataset was temporally split as follows: \emph{Train} — Jan 2012--Dec 2022; Jul 1--20 2023, \emph{Validation} — Jan--Jun 2023; Jul 22--30 2023, \emph{Test} — Aug 2023--Sep 2025.} 
\label{table:data}                             
\renewcommand{\arraystretch}{1.1} 
\begin{tabular}{|c||c c c|}        
\hline              
SXR Strength / Flare Class & Train & Validation & Test \\  
\hline
\quad $<$C & 11,712 & 257 & 13,600 \\
\quad C & 47,824 & 14,553 & 63,038 \\
\quad M & 16,469 & 1,487 & 22,368 \\
\quad X & 1,804 & 20 & 1,086 \\
\hline                             
\end{tabular}
\end{table}

\section{Methodology}
\label{sec:methodology}
\emph{FOXES}\footnote{See\dataset[doi: 10.57967/hf/8234]{https://huggingface.co/griffingoodwin04/FOXES} for access to the code base.} \citep{griffin_goodwin_2026_foxes} is a machine learning model specifically designed to map EUV data to GOES SXR measurements. Consequently, \emph{FOXES} functions as a translation tool rather than a forecasting or predictive model. For a given stack of EUV images (94\AA, 131\AA, 171\AA, 193\AA, 211\AA, 304\AA, and 335\AA) at a specified point-in-time, \emph{FOXES} outputs the corresponding GOES SXR integrated flux, as well as estimated SXR flux distributions over predefined patches on the solar surface. For this task, we settled on using a state-of-the-art deep learning framework known as Vision Transformers (ViTs). ViTs have demonstrated exceptional performance on a wide range of computer vision tasks \citep{vaswani2017,khan_transformers_2022, kameswari_overview_2023}, with recent heliophysics foundation models, such as Surya, using them as key components in their pipeline \citep{roy_surya_2025}. This success, combined with a ViT's ability to provide interpretable results, made it the best algorithm for this task. Please refer to Appendix \ref{sec:what_are_vits} to learn more about ViTs.

\subsection{The FOXES Architecture}
\label{sec:FOXES_arch}
\begin{figure}[h] 
  \centering
  \includegraphics[width=1\textwidth]{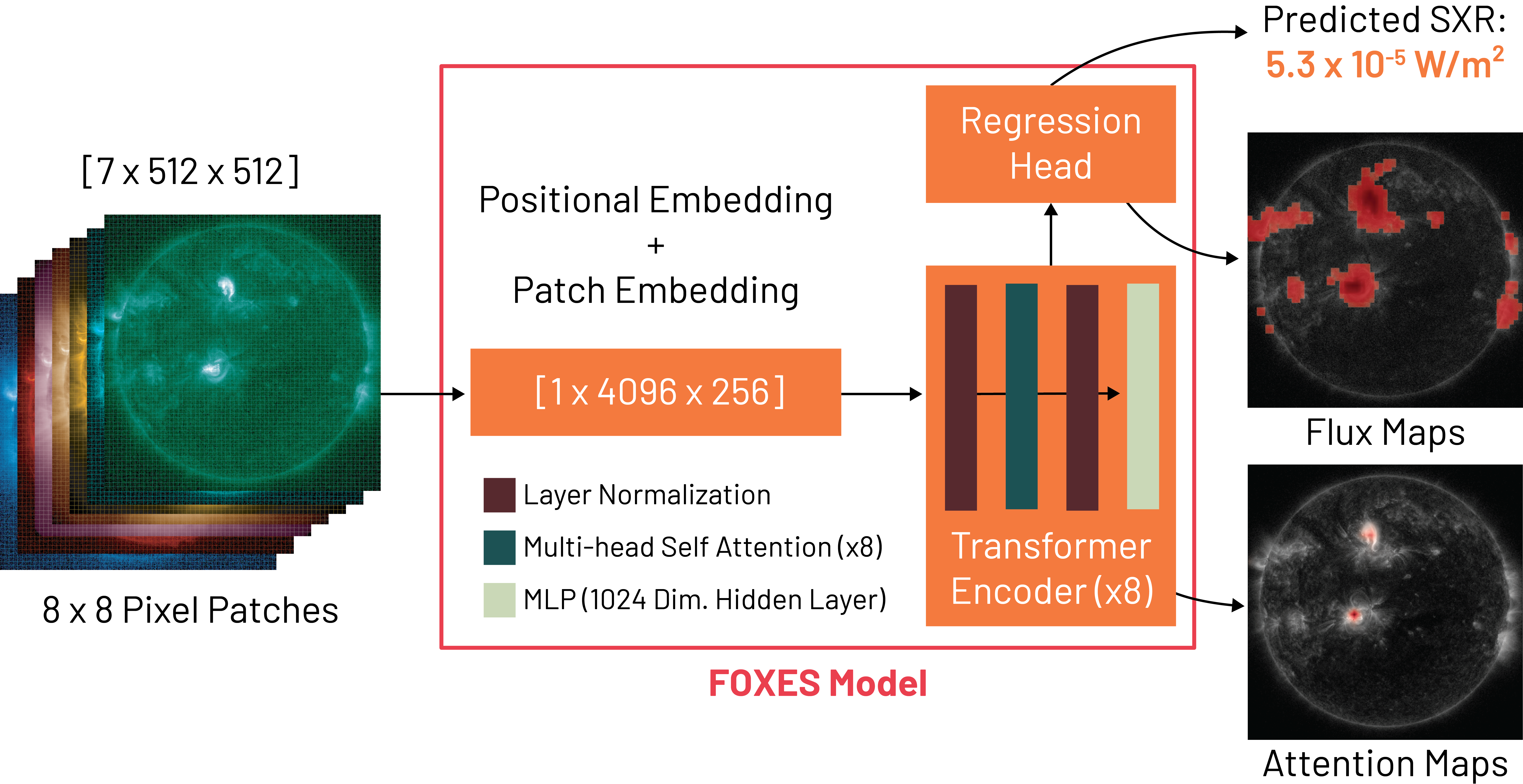}
  \caption{An overview of the \emph{FOXES} vision transformer architecture. A [7 $\times$ 512 $\times$ 512] array representing a stack of EUV images at a given time is first segmented into 8 $\times$ 8 pixel patches. These patches and their positions are flattened and fed into the transformer encoder block. Within this block exist layers for normalization, self-attention, and a traditional multilayer perceptron (MLP). The weights are then passed to a regression head for translation. There are three key outputs of \emph{FOXES}: the integrated GOES SXR translation, the spatially-distributed SXR flux maps, and the attention maps.}
  \label{fig:FOXES_arch}
\end{figure}
The ViT architecture behind \emph{FOXES} is based on a modified version of PyTorch Lightning's \citep{lightningai_vit_2025} code base. In the following section, we will give a brief description of the architecture and the changes we made to adapt this architecture to our task (see Figure \ref{fig:FOXES_arch}). \emph{Design choices were selected through iterative empirical evaluation. Further improvements to the model can likely be made through a thorough ablation study.}

As introduced above, the input to FOXES consists of a single stack of AIA images acquired at seven wavelength channels for a given observation time. Consequently, the model receives as input a three-dimensional array with dimensions [7 $\times$ 512 $\times$ 512]. Through experimentation, we tried a variety of patch sizes to strike a balance between performance and computational efficiency. Ultimately, we settled on a patch size of 8 $\times$ 8 pixels for a total of 64 $\times$ 64 or 4096 patches per image. These patches are then flattened into an array of 448 values (7 wavelengths $\times$ 8 pixels $\times$ 8 pixels) and fed into our linear projection layer. We selected an embedding dimension of 256, converting the input dimension into a three-dimensional array of  [1 $\times$ 4096 $\times$ 256]. These patch embeddings are then fed into our transformer encoder block, which consists of layer normalizations, eight multi-head self-attention mechanisms, and a 1024-dimensional MLP. We employ specialized masked-attention layers to emphasize global context, while suppressing redundant local information. This architecture encourages the model to focus on two complementary signals. First, the attention mechanism learns how each patch compares to the global background state of the Sun. Second, the residual connections preserve the local brightness and structure of each patch. This enables the model to make robust patch-based translations. We selected a masking grid of 9 $\times$ 9 patches surrounding the reference patch, meaning any given patch can only communicate with patches outside this localized 9 $\times$ 9 grid. The transformer encoder block was repeated eight times and then fed into a regression head. This regression head then applies a linear activation function to make a prediction for the amount of SXR flux coming from each 8 $\times$ 8 pixel patch fed into the model. This essentially gives us a virtual SXR flux map that we can use to localize the integrated flux (see the outputs in Figure \ref{fig:FOXES_arch}). In addition, because \emph{FOXES} is derived from a vision transformer architecture, we can extract the attention maps for the given AIA image stack that we provide from the transformer encoder block (see the outputs in Figure \ref{fig:FOXES_arch}). This essentially gives us a sanity check to assess whether the model produces physically meaningful translations. In particular, it enables us to verify that the model is focusing on relevant features in the images (e.g., active regions, filaments, etc.) rather than relying on spurious correlations or random patterns present in the data to generate its predictions.

During model training, we employed a batch size of 12, which we identified as optimal for training on two NVIDIA A100 GPUs. The model was trained for 100 epochs, while employing a cosine learning-rate scheduler that decays from $10^{-4}$ to $\sim10^{-5}$. 

Model loss was computed in a normalized log space (based on the training dataset) using the Huber loss function, defined as
\begin{align}
\mathcal{L}_\delta =
\begin{cases}
\frac{1}{2}(F_i-\hat{F}_i)^2, & |F_i-\hat{F}_i|\le \delta \\
\delta|F_i-\hat{F}_i|-\frac{1}{2}\delta^2, & \text{otherwise}
\end{cases}
,\quad \text{with } \delta = 0.3.
\end{align}
Here, $F_{i}$ denotes the observed GOES SXR flux for a sample $i$, and $\hat{F}_{i}$ represents the predicted integrated SXR flux of the model, obtained by summing the fluxes in each patch. The parameter $\delta$ is defined as a threshold value, which controls the transition between the mean absolute error and mean squared error loss. If the residual error is small ($|F_i-\hat{F}_i|\le \delta$), the mean squared error is used; otherwise, the mean absolute error is used. We found that a $\delta = 0.3$ provided reasonable results. 

Finally, the loss was weighted by the inverse frequency of each flare class (flare quiet, C, M, X) in the training dataset, a common technique used in other rare-event forecasting works \citep{ahmadzadeh_how_2021}. The weights were adapted periodically during training depending on the historical performance of each class. This ensured that class imbalance was mitigated during training.

\subsection{Baseline}
\label{sec:methods_baseline}
To quantify the performance of \emph{FOXES}, we also developed a simple baseline to use for comparison. Specifically, we employed a linear regression model with the same translational functionality as \emph{FOXES} that utilizes the disk-integrated signal in 7 EUV channels and maps it to the 1--8\,$\rm \AA$ SXR flux. By design, this model does not account for the spatial locations of the pixels, and thus it is unable to distribute the integrated flux across the solar disk. This enabled us to get a basic understanding of how spatial information impacted our results. To deploy such a model required a slight modification to our dataset. Instead of using the spatially-resolved image stack (7 $\times$ 512 $\times$ 512), we created an EUV spectral irradiance index  by summing the pixel intensities in each channel (7 $\times$ 1 vector).

\subsection{Performance Metrics}
\label{sec:performance_metrics}
We evaluated our baseline and \emph{FOXES} models using several regression metrics, including mean absolute error (MAE), root mean squared error (RMSE), mean bias error (MBE), and Pearson correlation coefficient (r), comparing the total summed patch flux with the ground truth GOES integrated flux. These metrics provide complementary characterizations of performance by probing distinct error regimes. The MAE quantifies the typical magnitude of prediction errors and offers a robust measure of central tendency that is largely insensitive to outliers, while RMSE imposes progressively stronger penalties on large deviations, thereby emphasizing the model behavior during sporadic high-flux intervals, including intense flaring events. The MBE provides some insight into the model tendency to over/underpredict certain flare classes by calculating the average signed residual between the predicted and ground truth values. Lastly, the Pearson correlation coefficient allows us to quantify the strength of the linear relationship between the predicted and ground truth flux. This can help us identify potential model discrepancies that may not be obvious in the scalar errors. Because the SXR irradiance spans several orders of magnitude, all errors were computed in $log_{10}$ space.

\section{Results}
\label{sec:results}
Our results are organized as follows. Section \ref{sec:baseline} establishes the results of our baseline model. Section \ref{sec:FOXES_res} describes the performance provided by the \emph{FOXES} model for translating EUV images to integrated GOES SXR flux and how it improves upon the baseline. Finally, in Section \ref{sec:potential_implications}, we describe potential future applications of \emph{FOXES}.

\subsection{Baseline EUV to GOES Integrated SXR Flux}
\label{sec:baseline}
The results of our baseline model are plotted in the top panel of Figure \ref{fig:regression}. It shows a 2D hexagonal binned plot comparing the translated flux by the model on the y-axis to the ground truth GOES SXR flux on the x-axis. A few details stand out. It is evident that the baseline model performs particularly poorly at both lower- and higher-end fluxes. There are several potential explanations for this behavior. First, no sample weighting scheme was found to improve performance, so the model was trained without one. Consequently, because the majority of the training samples are at a C-class flux, the model is biased toward predicting C-class output. This results in an overestimation of events with fluxes weaker than C-class and a frequent underestimation of events stronger than C-class, something which is consistent with our observations. Additionally, we see that the baseline tends to overestimate the strength of M- and X-class fluxes. This may arise from the saturation of AIA pixels and diffraction effects from the entrance grid in bright events, which affect the brightness of areas close to the flare. This, in turn, leads to overestimation of the total EUV emission when the data are spatially integrated. This is an area where \emph{FOXES} can improve translations. By including the spatial information during training, the model can learn to take into account the shape and extent of the oversaturated pixels.

In Table \ref{table:qualitymetrics}, we quantify the performance of our linear regression baseline model using the metrics discussed in Section \ref{sec:performance_metrics}. The results are in agreement with the discussion above, with performance across all four metrics peaking at a C-class level. Overall, the model provides a translation MAE of 0.307 dex. This means that for any given flare class, on average, our predictions are off by a factor of $10^{0.307}\approx2.03$. For example, a true C5.0 class GOES flare would have an error between $\sim$C2.5 and $\sim$M1.0 \emph{FOXES} magnitude. This isn’t the worst, but it’s definitely less than ideal.

\subsection{FOXES EUV to GOES Integrated SXR Flux}
\label{sec:FOXES_res}
\begin{figure}[p]
  \centering
\includegraphics[width=1\textwidth]{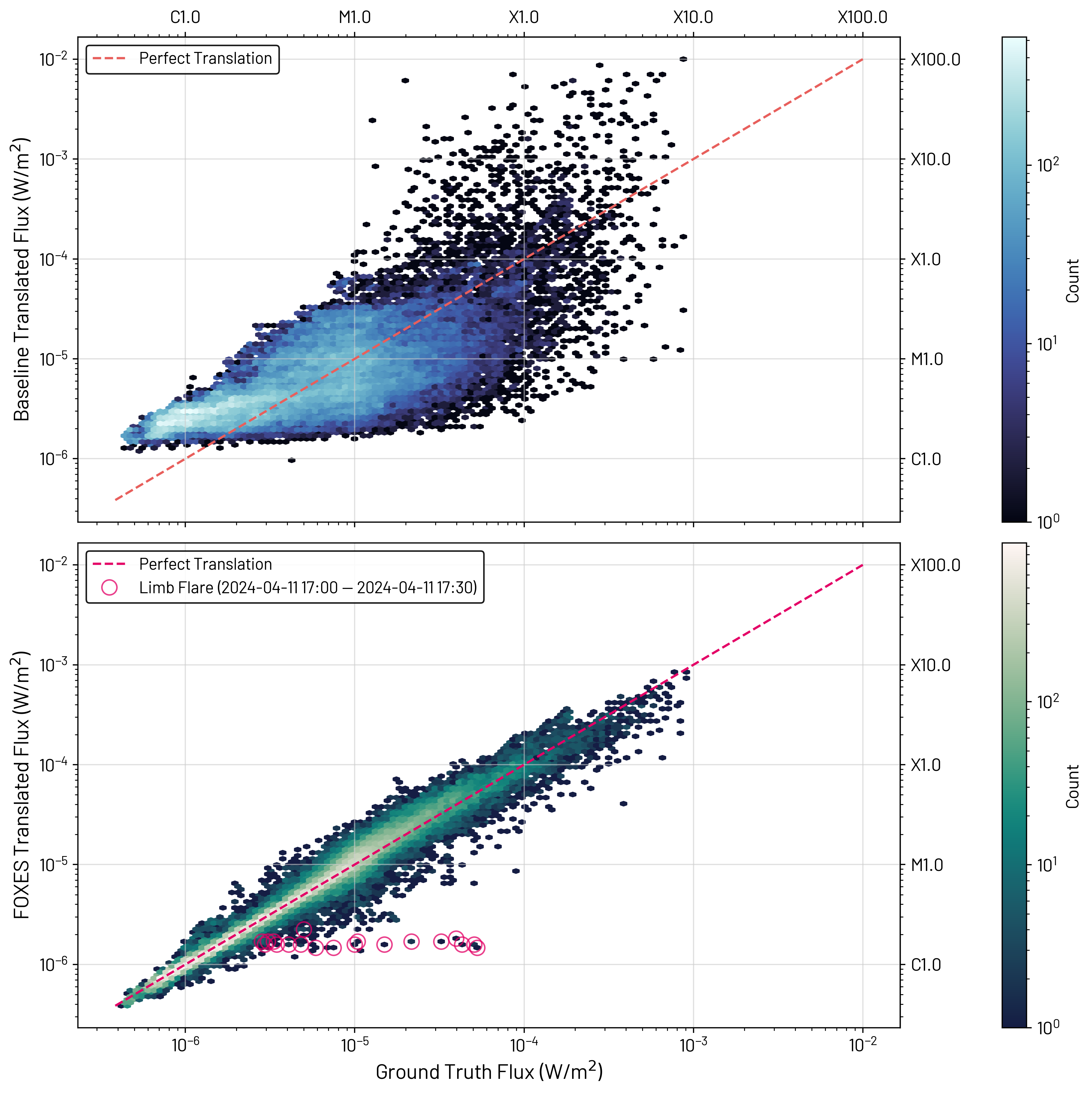}
  \caption{\emph{(Top Panel)} A 2D hexagonal binned plot comparing the baseline model's integrated translations to the GOES integrated SXR flux (ground truth). A perfect translation (red line) is overlaid on top. \emph{(Bottom Panel)} Same for the \emph{FOXES} model's integrated translations (by summing up the flux from each patch) to the GOES integrated SXR flux (ground truth). A perfect translation line is also overlaid on top. The bins corresponding to a limb flare that occurred on April 11th, 2024, are highlighted in this panel by pink circles and discussed further within the text.}
  \label{fig:regression}
\end{figure}

  We repeat the analysis in the previous Section, now comparing the predicted SXR flux from the individual patch-wise translations produced by \emph{FOXES} with the corresponding GOES flux for the entire testing dataset. The bottom panel of Figure \ref{fig:regression} summarizes our results. Compared to the top baseline panel, it is evident that our results have improved significantly by using the more sophisticated ViT model, in conjunction with the learned spatial information. There is obviously lower bias, with predictions hovering \emph{evenly} around the perfect correlation line for nearly every flare class. Only at the highest fluxes (X+) does the model begin to underpredict noticeably. We interpret this as a consequence of the insensitivity of the AIA filters to plasmas with temperatures $\gtrsim30$ MK produced in these flares \citep{caspi_statistical_2014}. The peak temperature response of AIA’s hottest channel, 193\AA, occurs at approximately 20 MK \citep{Lemen2011}. This implies that the hottest plasmas, which contributes substantially to the emission observed in the GOES SXR 1--8$\rm \AA$ channel during intense X-class solar flares, is largely invisible to AIA.


\begin{table}[h]
\caption{Performance metrics calculated in log-space for the baseline model and \emph{FOXES}. Percent improvements are shown in parentheses for MAE and RMSE ($\downarrow$ = error reduction).}
\label{table:qualitymetrics}
\centering 
\renewcommand{\arraystretch}{1.2}
\begin{tabular}{|c||c|cccc|}
\hline
\textbf{Model} & \textbf{Class} & \textbf{RMSE} & \textbf{MAE} & \textbf{MBE} & \textbf{r} \\
\hline\hline
\multirow{5}{*}{\rotatebox{90}{\textbf{Baseline}}} 
& \textbf{Overall} & \textbf{0.376} & \textbf{0.307} & \textbf{0.116} & \textbf{0.737} \\
\cline{2-6}
& $<$C & 0.474 & 0.468 & 0.468 & 0.571 \\
& C & 0.296 & 0.250 & 0.163 & 0.636 \\
& M & 0.436 & 0.349 & $-$0.225 & 0.419 \\
& X & 1.090 & 0.718 & $-$0.010 & 0.400 \\
\hline\hline
\multirow{5}{*}{\rotatebox{90}{\textbf{\emph{FOXES}}}} 
& \textbf{Overall} & \textbf{0.079} \textcolor{ForestGreen}{\textbf{(79\%$\downarrow$)}} & \textbf{0.051} \textcolor{ForestGreen}{\textbf{(83\%$\downarrow$)}} & \textbf{0.083} & \textbf{0.990} \\
\cline{2-6}
& $<$C & 0.033 \textcolor{ForestGreen}{(93\%$\downarrow$)} & 0.026 \textcolor{ForestGreen}{(94\%$\downarrow$)} & $-$0.005 & 0.919 \\
& C & 0.063 \textcolor{ForestGreen}{(79\%$\downarrow$)} & 0.042 \textcolor{ForestGreen}{(83\%$\downarrow$)} & 0.003 & 0.980 \\
& M & 0.121 \textcolor{ForestGreen}{(72\%$\downarrow$)} & 0.088 \textcolor{ForestGreen}{(75\%$\downarrow$)} & 0.035 & 0.901 \\
& X & 0.169 \textcolor{ForestGreen}{(85\%$\downarrow$)} & 0.130 \textcolor{ForestGreen}{(82\%$\downarrow$)} & $-$0.057 & 0.710 \\
\hline
\end{tabular}
\end{table}

Table~\ref{table:qualitymetrics} summarizes the results for both the \emph{FOXES} and the baseline model. As shown in the parenthesized values, \emph{FOXES} greatly improves upon the simple linear regression model in every performance metric tested. \emph{FOXES} provides an overall translation MAE of 0.051 dex, which corresponds to an average factor of  $10^{0.051}\approx1.12$. For example, for a GOES C5.0 flare, the \emph{FOXES} prediction lies between  $\sim$C4.5 - 5.6, an impressive result. The MBE values indicate that the model continues to underpredict in both ends ($<$C and X-class), but these biases are much reduced over the baseline (aside from those at the X-class level).

\begin{figure}
  \centering
    \includegraphics[width=1\textwidth]{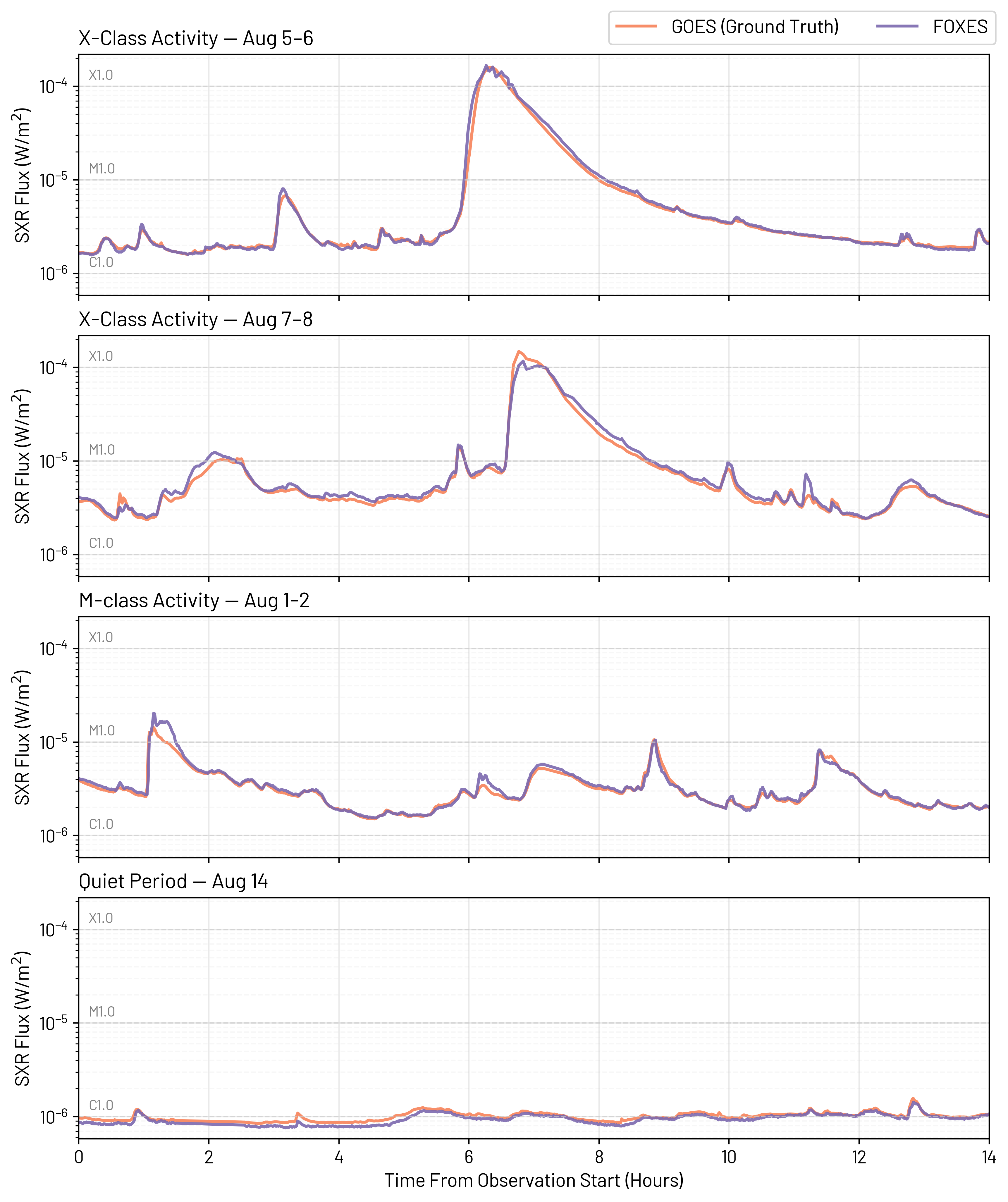}
  \caption{An application of \emph{FOXES} to AIA data sampled with 1-minute cadence across 4 time periods in August 2023. Each window is 14 hours long. The orange line highlights the ground truth SXR flux from GOES, while the purple line is the \emph{FOXES} translation. \emph{Note: These translations are still being made at the given observation time. We are not forecasting ahead or using previous data to inform the model.}}
  \label{fig:timeseries}
\end{figure}

To further demonstrate the capability of \emph{FOXES}, we tested it on continuous AIA data sampled with 1-minute cadence for several flaring events during August 2023 (see Figure \ref{fig:timeseries}). Note that \emph{FOXES} uses point-in-time data for the translations and has no temporal context of the data that came before the observed timestamp. Despite this limitation, \emph{FOXES} was still able to capture not only the growth and peak phases of these events, but also their decay phase. This indicates that 1-minute EUV data can capture the rapid heating of the plasmas in the flare site detected by GOES. 

\begin{figure}
  \centering
    \includegraphics[width=0.75\textwidth]{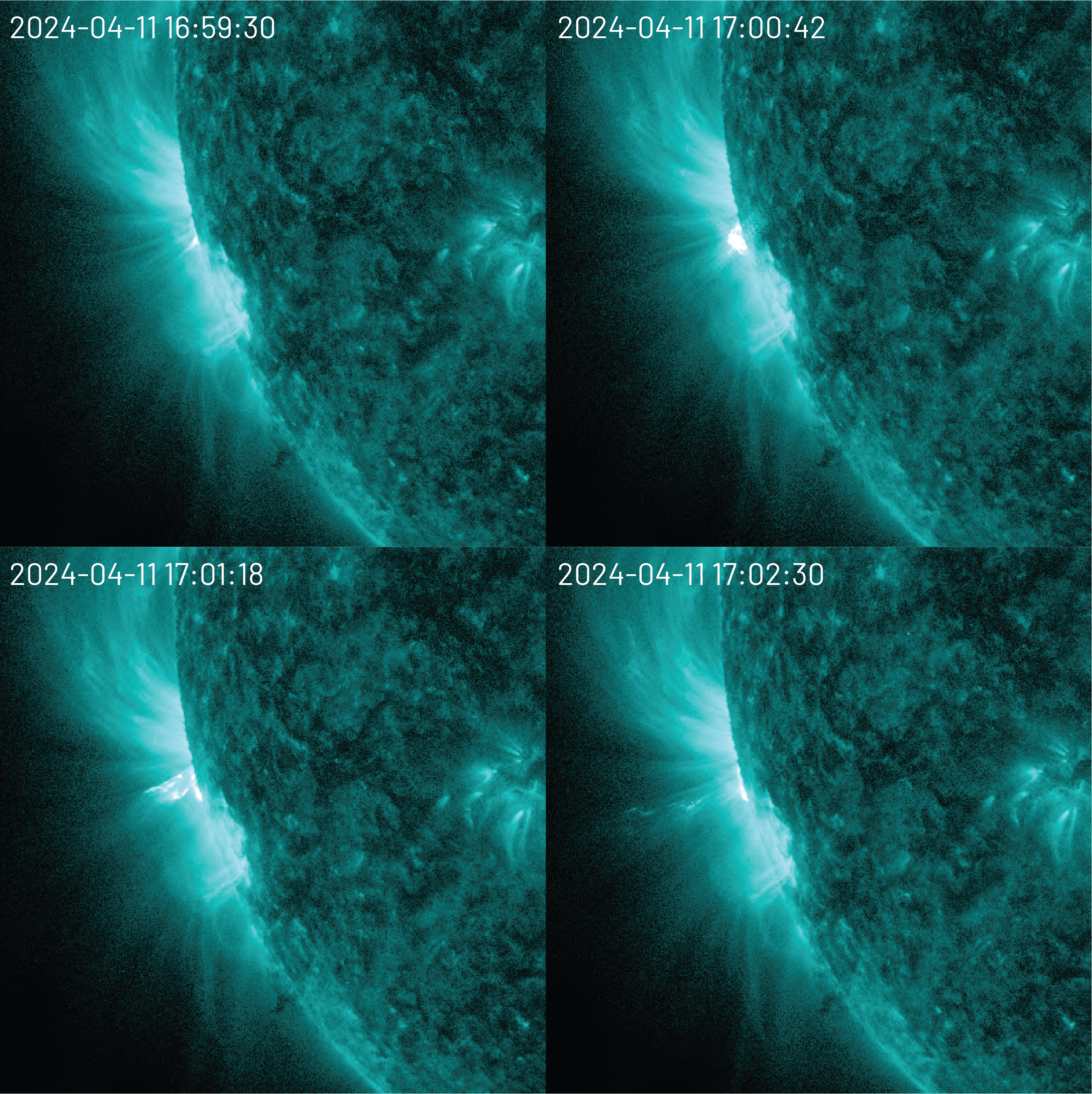}
  \caption{SDO/AIA 131\AA~ snapshots of the M-class flare that occurred on April 11th, 2024.} 
  \label{fig:limbflare}
\end{figure}

Finally, we briefly investigate the most obvious outliers in the bottom panel of Figure~\ref{fig:regression}. Near the M-class ground truth and C2.0 prediction corner, a clear streak of points protrudes from the continuous cluster (see the pink circles). These points correspond to observations on 11 April 2024. Figure \ref{fig:limbflare} shows snapshots of the flaring source on this day. It is a limb flare and may explain why our model underpredicted this particular event. EUV flare loops near the limb may be optically thick \citep{thiemann_center--limb_2018}, which could lead to lower EUV intensities compared to SXR emission. Moreover, we crop the off-limb corona to 1.1 solar radii. This operation can potentially cut off excess emission that may have been captured by GOES. 
To further investigate this behavior, we evaluate the ``spatially distributed flux-weighted performance" of \emph{FOXES}. For each timestamp within the test dataset, we first extract the corresponding flux maps and normalize them such that they sum to one. We then multiply all individual flux patches by the error of the corresponding timestamp (the MAE / MBE of the predicted and ground truth integrated SXR flux in log$_{10}$ space). By summing these scaled maps and dividing by the total flux for each flux patch over all timestamps, we obtain a map that can be interpreted as the typical flux-weighted error observed at a given ViT patch. This error can be mathematically represented by Equation \ref{eq:MAE_MBE}: 
\begin{equation} \label{eq:MAE_MBE}
    \text{Flux-Weighted Error}(i,j) = \frac{\sum_{t} F_{ij} \times \delta_t}{\sum_t F_{ij}}, \quad
    \delta_t = 
    \begin{cases}
        |\log_{10}(\hat{y}_t) - \log_{10}(y_t)| & \text{(MAE)} \\
        \phantom{|}\log_{10}(\hat{y}_t) - \log_{10}(y_t)\phantom{|}  & \text{(MBE)}
    \end{cases}
    .
\end{equation}
Here, $F_{ij}$ is the flux at a given ViT patch $(i,j)$, $\hat{y}_t$ is the predicted integrated flux at time $t$, and $y_t$ is the ground-truth integrated flux at time $t$. If we observe a clear distribution of the flux-weighted error toward the solar limb, we can obtain indirect evidence that \emph{FOXES} tends to provide more inaccurate integrated SXR flux translations during limb events.

\begin{figure}
  \centering
    \includegraphics[width=1\textwidth]{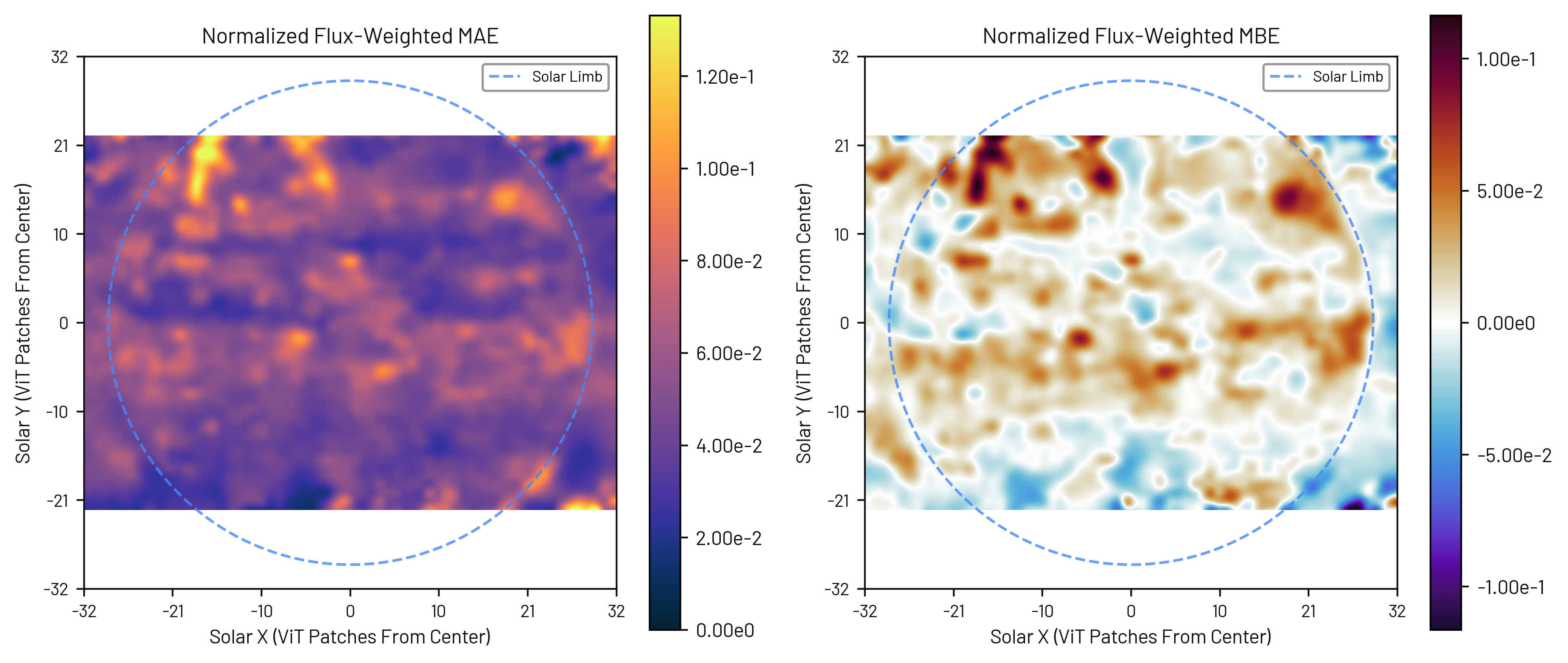}
  \caption{Normalized flux-weighted maps for mean absolute error (\emph{Left}) and mean bias error (\emph{Right}). Each map was calculated using the equation defined in \ref{eq:MAE_MBE}. Patches above and below $\approx \pm$56$^\circ$ in disk latitude were removed as they are not relevant to this investigation, and tend to be noisy due to their low flux values. These maps can be interpreted as how the integrated SXR error of \emph{FOXES} varies with the predicted spatial distribution of the flux.} 
  \label{fig:performance_heatmap}
\end{figure}

Figure \ref{fig:performance_heatmap} illustrates the results of our investigation. What is evident from both plots is that there is no clear bias towards disk-center or limb events, with the MAE relatively evenly spread across the map. Looking at the MBE, we observe a slight asymmetry, with \emph{FOXES} tending to overpredict on the eastern limb and underpredict on the western. However, this may reflect a bias within the samples themselves rather than a systematic bias of \emph{FOXES}. Overall, the results suggest that our previous intuition is plausible. The event from Figure \ref{fig:limbflare} was likely anomalous and either developing within large coronal loops (which were omitted in EUV images due to 1.1 solar radii cropping) or potentially accompanied by an optically thick EUV emission (due to foreshortening effects). It would be valuable to investigate this further in future studies, where a new \emph{FOXES} model can be trained on uncropped images.

\subsection{Potential Implications}
\label{sec:potential_implications}
\begin{figure}[] 
  \centering
    \subfloat[][]{\includegraphics[width=1\textwidth]{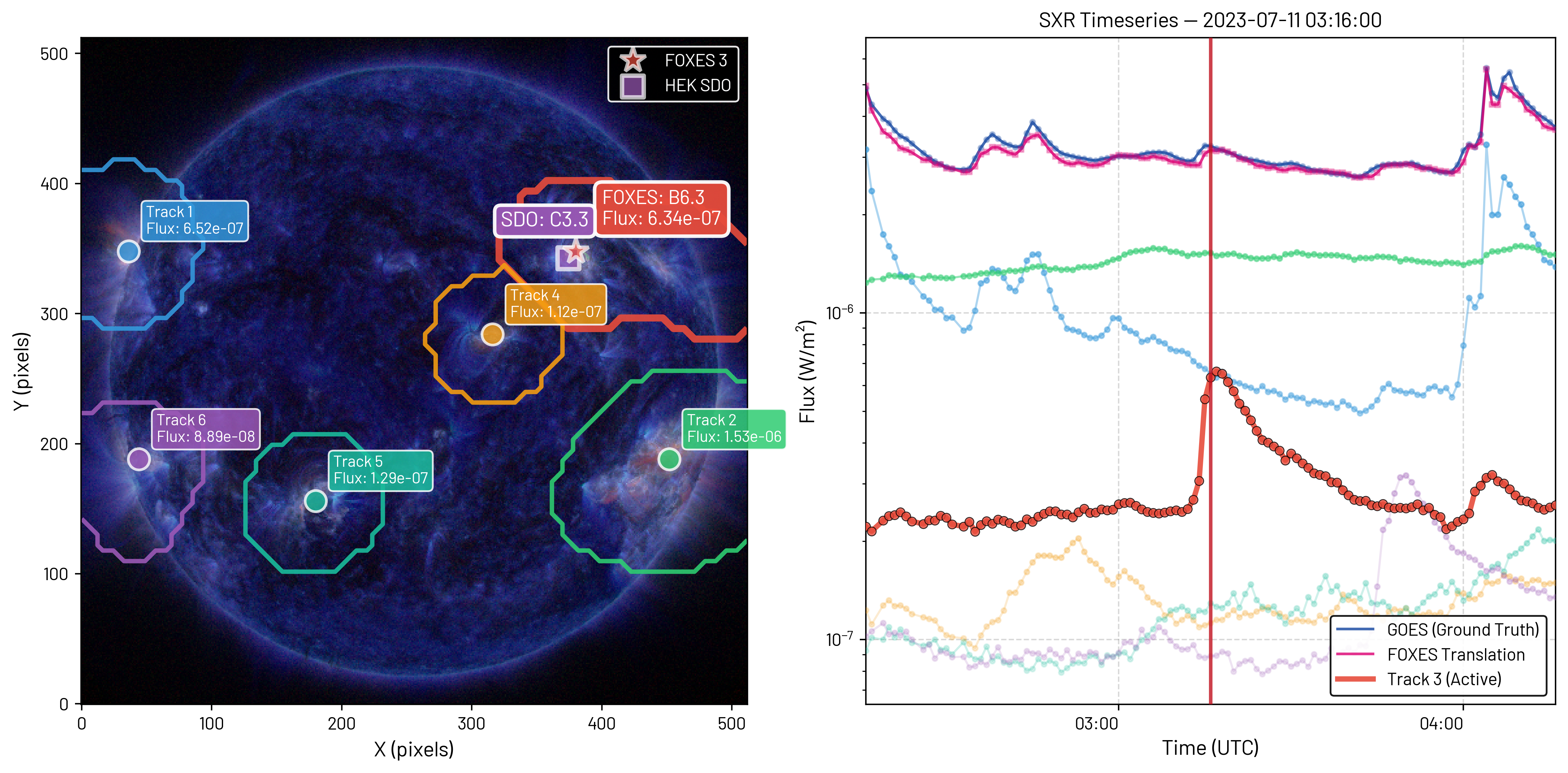}}\\
    \vspace{1em}
    \subfloat[][]{\includegraphics[width=1\textwidth]{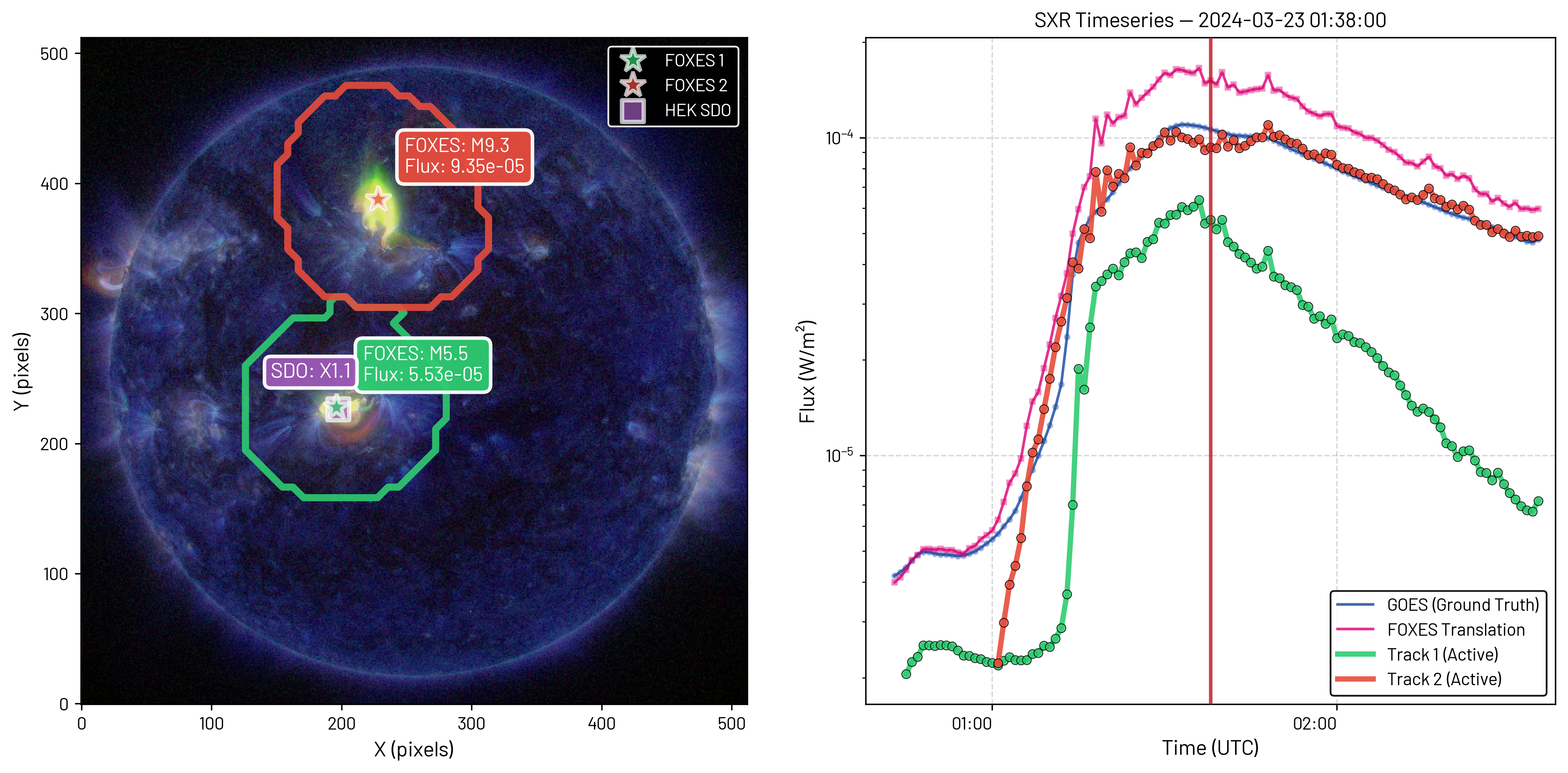}}
  \caption{On the left of both (a) and (b), we display a stack of AIA images (131\AA, 171\AA, 304\AA) at a given time, along with the active regions (high flux regions) identified by \emph{FOXES}. We mark flaring locations identified by \emph{FOXES} (stars) and by the Heliophysics Event Knowledge database's SDO/AIA detection algorithm (squares). The integrated flux of the patches in each individual region track is shown as well. On the right, we display the corresponding soft X-ray translations for each region over a $\pm$1-hour window. The dark-blue curve at the top is the integrated GOES ground truth, and the pink curve on the top is the integrated \emph{FOXES} curve. Everything below are individual region tracks color-coded in agreement with the left panel. \emph{Please note that the example shown in (a) was selected from the training dataset. This should not cause any issues or introduce bias, as we are simply trying to explore \emph{FOXES'} ability to distribute the SXR flux.}}
  \label{fig:sxr_disection}
\end{figure}

Motivated by the promising results of \emph{FOXES}, we have considered several options for future improvements and downstream applications. Since \emph{FOXES} not only translates EUV images into \emph{integrated} GOES SXR flux, but also distributes this flux across the solar surface, we could use \emph{FOXES} to produce a catalog of active region (AR)-based flare events and determine the SXR classes more accurately. An example of what this may look like is shown in Figure \ref{fig:sxr_disection}. We first extract the 8~x~8 pixel patch flux maps provided by \emph{FOXES}. For each map, we then apply a Gaussian blurring to ensure that captured high intensity regions remain consistent across time stamps because ARs regularly change shape and size during their evolution. The Gaussian-blurred maps are then segmented into ARs based on local maxima observed within the map. In Figure \ref{fig:sxr_disection}, the sum of the patches for each region is depicted with a different color, along with the HEK flare record and \emph{FOXES}-derived GOES class. In the right panel, we compare the integrated GOES and \emph{FOXES} intensities (top dark blue and pink curves, respectively). The other curves show the individual AR light curves, as derived from \emph{FOXES}. The color of each curve corresponds to the color of the AR patch in the left panel. In (a), we see that the observed C3.3 flare is correctly attributed to the AR in the northwest corner, but \emph{FOXES} identifies it as a weaker, B6.3 class, flare with respect to what is recorded by GOES. Given that the light curve appears physically plausible, this is a promising result: \emph{FOXES} seems to be able to separate each AR’s contribution from the global background. A similar example is presented in panel (b), which shows two simultaneously-ongoing (i.e., sympathetic) flares. According to the HEK catalogue, an X-class flare is identified only in the lower active region, with no additional events reported (the AIA flare trigger module does identify the northern region, but no SXR flux is attributed to it). In contrast, \emph{FOXES} correctly identifies the presence of two distinct events ($\sim$ higher-M-class) and, moreover, indicates that the northernmost flare is likely more intense than the southern one. This distinction in labeling is critical, especially for flare forecasting works. While \emph{FOXES} slightly overpredicts the integrated flux in this case, this example nonetheless further demonstrates the potential of the model.  

Of course, these are just proof-of-concept examples, which require additional testing. Validation is challenging since there are no AR-isolated flare classes observed. Such work will require (1) checking against Hinode XRT images when available and (2) comparing with flares when only a single AR exists on the disk. It may also be of interest to compare our results to those by \cite{padial_doble_alexis_2025}, who provide a collection of flaring events with spatially dissected integrated SXR flux. However, the overlap between their high-cadence data and ours is currently limited, and a rigorous comparison would require substantial additional data preprocessing. We leave this to future work. 

\begin{figure}[]
  \centering
    \includegraphics[width=1\textwidth]{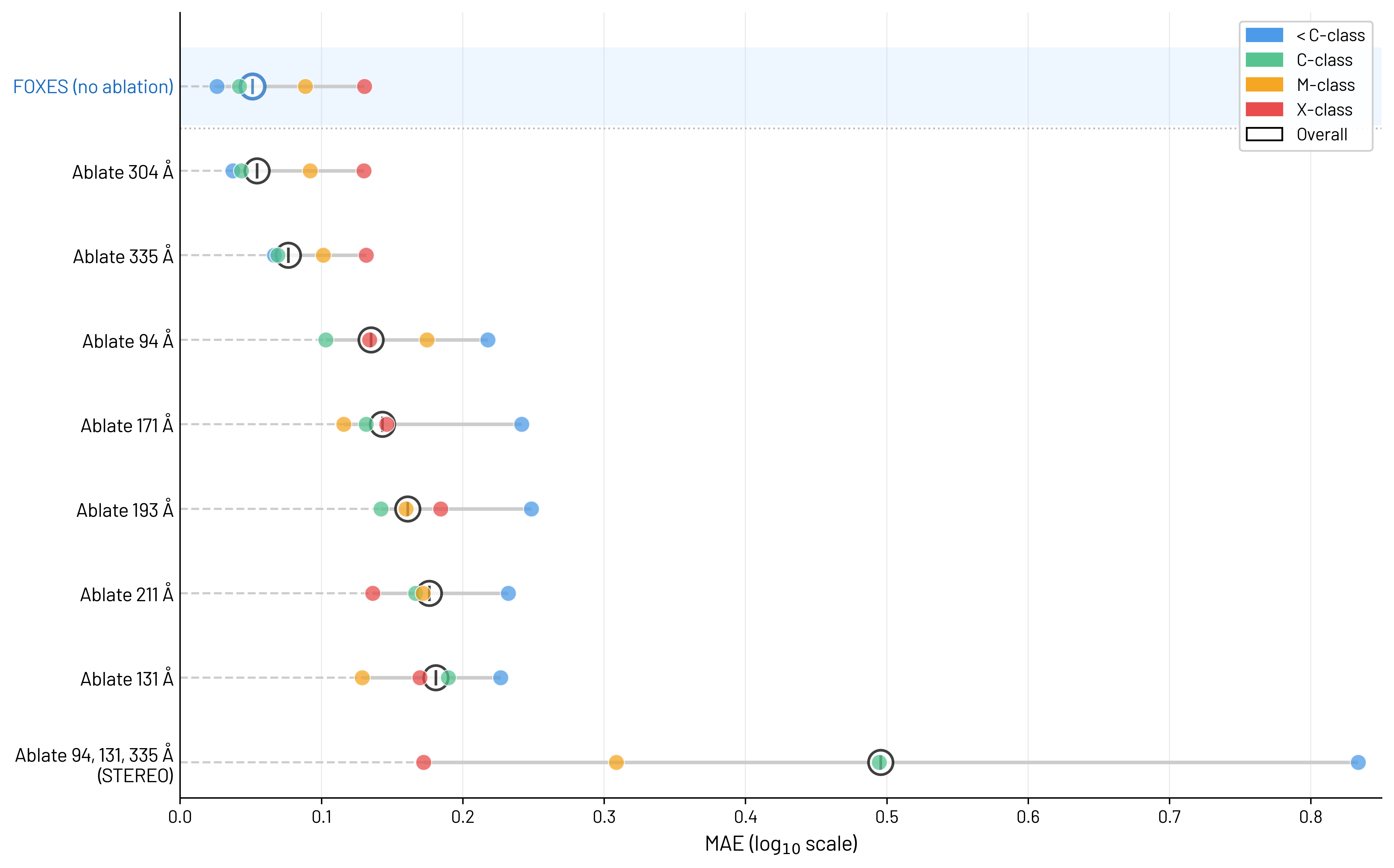}
  \caption{The mean absolute error (calculated in log$_{10}$ space) across flare classes (and overall) for \emph{FOXES} applied to ablated versions of the test dataset, where individual EUV channels (and one triple-channel combination) were augmented with Gaussian noise.} 
  \label{fig:lollipop}
\end{figure}

\emph{FOXES} is poised to make a unique contribution to the large number of EUV observations from missions outside the Sun-Earth line (SEL), such as the currently operating STEREO and Solar Orbiter missions and the planned ESA Vigil mission. None of these missions carry an SXR imager, but they all carry multi-channel EUV imagers. STEREO, in particular, has recorded a larger number of flares, either at the limb or on the far side of the Sun. However, it is difficult to relate the strength of these flares to the historical data observed by GOES and to apply/validate any flare-based forecasting schemes. By providing a `virtual GOES', \emph{FOXES} will help close this important gap. Within the ITI framework, \emph{FOXES} could in principle be extended to STEREO/EUVI and Solar Orbiter/EUI observations \citep{jarolim2025, schirninger2025iti}. However, with fewer EUV channels (four for STEREO, only two for Solar Orbiter), it seems likely that the SXR predictions will not be as robust as those derived from the seven SDO channels. Here we provide a brief investigation into how the performance of \emph{FOXES} may be impacted by a reduced number of input channels. Given the complexity of the \emph{FOXES} model and associated computational training costs, a proper ablation study is not feasible at this point and will be a subject of future studies (given the computational resources are available). Instead, our approach is as follows: for each wavelength, as well as the trio of channels \textit{unavailable} to STEREO (94, 131, 335\AA), we apply Gaussian noise (mean of 0 and standard deviation of 1) to the images in the test set, clamping the images to [-1,1] (the range of the normalized inputs). This results in eight (7 individual wavelengths and 1 triple-channel ablation) new test datasets that we can run inference on using the original \emph{FOXES} model. Our results are shown in Figure \ref{fig:lollipop}, which clearly highlight the most influential (131\AA, 211\AA, 193\AA) and least impactful (304\AA, 335\AA) wavelengths for translations. These results tell an expected story: wavelengths with response functions that peak near or cover a large portion of $\geq$C-class flaring temperatures ($\geq$5MK) tend to be the most useful, while wavelengths that peak near cooler temperatures are not. Interestingly, though, the 94\AA~channel, which also peaks at these high temperatures, seems to have a minimal impact on performance when ablated. One possible reason for this outcome is that \emph{FOXES} learned to pay less attention to this channel due to its relatively high noise level. The model likely realized that similar information was available from other channels that were less contaminated by noise. What we also see from the results is the severe impact on performance when ablating the 94\AA, 131\AA, 335\AA~channels together. There is a major dip, with the overall MAE hovering close to 0.49 dex. In the future, we plan to perform a full retraining of the model using the STEREO wavelengths to get a better understanding of the limitations we observe here. However, if we can produce a reliable version of the model with these reduced wavelengths, the expansion of \emph{FOXES} to STEREO, and potentially to Solar Orbiter and eventually Vigil, will greatly enhance our ability to support human exploration of Mars by providing widely-used soft X-ray solar activity observables using existing instruments.

\section{Conclusion}
\label{sec:conclusions}
\emph{FOXES} demonstrates that multichannel EUV images can be used to reasonably estimate the integrated GOES SXR flux, with an overall MAE of 0.051 dex. The possible exceptions to this are limb flares, where further investigation and model optimization are necessary. In addition, from our brief case test, \emph{FOXES} appears to be able to help locate the sources of SXR emission and thus `calibrate' the GOES SXR curves and flare classes, particularly at times of high activity.

The preliminary success of this proof-of-concept opens several exciting avenues of research. For missions outside the Sun-Earth line, such as STEREO (and possibly Solar Orbiter), \emph{FOXES} could transform their multichannel EUV images into a `virtual' GOES, allowing the extraction of SXR light profiles and the classification of flare strength consistently around the Sun. This not only extends the historical record of flares but also enables quantitative comparisons of far-side to front-side events. 
From a space weather perspective, \emph{FOXES} enables the deployment of flare-based forecasting approaches to off-Sun-Earth line vantage points, thus providing support for human exploration to Mars. On the science side, \emph{FOXES} shows promising results for predicting SXR flux across individual sub-regions, enabling localization and characterization of simultaneous flares (though more testing is required) and a more-accurate estimate of the SXR fluxes from individual ARs. Overall, \emph{FOXES} has the potential to support the creation of a more comprehensive and reliable flare catalog, which could in turn further improve current flare forecasting algorithms and provide better statistics on flare occurrence. With the successful application of \emph{FOXES} in the future, these advancements would represent a significant step forward for the space weather community.

\section{Acknowledgments}
This work is a research product of Heliolab (heliolab.ai), an initiative of the Frontier Development Lab (FDL.ai). FDL is a public–private partnership between NASA, Trillium Technologies (trillium.tech), and commercial AI partners including Google Cloud and NVIDIA.
Heliolab was designed, delivered, and managed by Trillium Technologies Inc., a research and development company focused on intelligent systems and collaborative communities for Heliophysics, planetary stewardship and space exploration.
We gratefully acknowledge Google Cloud for extensive computational resources and NVIDIA Corporation.
This material is based upon work supported by NASA under award number No. 80GSFC23CA040. Any opinions, findings, conclusions, or recommendations expressed are those of the authors and do not necessarily reflect the views of the National Aeronautics and Space Administration. Griffin Goodwin would like to acknowledge NASA FINESST grant 80NSSC23K1639.

Large language models were used as brainstorming tools to discuss possible training strategies and methodological considerations. The authors retained full responsibility for all research decisions, interpretations, and conclusions presented in this work.

\section{Appendix}
\subsection{What Are Vision Transformers}
\label{sec:what_are_vits}
Vision Transformers operate analogously to how many large language models (LLMs) function. However, rather than receiving a tokenized text sequences as input, ViTs process images. At first glance, this may seem unintuitive, yet the connection is fairly straightforward. The input image (in our case, EUV observations) can be partitioned into a set of smaller, fixed-size sub-images, commonly referred to as patches. Each patch is treated as a discrete token, in the same way that individual words function as tokens within a sentence for LLMs. The ordered collection of all patches reconstructs the full ``sequence'' representation of the original image, giving the model something similar to a paragraph that it can ingest. 

ViTs consist of two key components: the transformer encoder block and the multilayer perceptron (MLP; simple feed-forward neural network) head. The guts of the model lie within the transformer encoder block, which does most of the heavy lifting. Once the patches have been defined, the pixels from each patch are flattened into a one-dimensional array. We then tack on an additional dimension onto these arrays, which acts as the positional encoding for the patch, allowing the model to learn where each patch originates in the image. For our particular model, we employ a two-dimensional-positional encoding. From here, the flattened patches pass through several self-attention mechanisms. In a single self-attention network, each patch array is delegated a learned query (Q), key (K), and value (V) representation. The query can be interpreted as what a given patch is looking for, the key describes what that patch contains, and the value is what the patch will share with the other patches. From there, the attention is calculated across the patches by taking the dot product of the query for a given patch and the keys of all other patches. The raw similarity scores are then normalized and passed through a soft-max function, which describes how strongly each patch influences one another. Finally, these weights are multiplied by the corresponding values of the appropriate patches, allowing the model to gain context from other patches.

Multi-head self-attention repeats this process in parallel several times to capture different spatial and structural aspects of the patches. The resulting outputs are then fed into an MLP within the encoder block, and the block is stacked multiple times. Finally, the encoded representation from the transformer is passed to a classification or regression head to provide a prediction.  
\bibliography{sample701}{}
\bibliographystyle{aasjournalv7}



\end{document}